\documentclass[12pt]{article}
\usepackage{graphicx,amsmath,amssymb,amsfonts,color}
\usepackage{citesort,epsfig}

\input{text/stability.pre}
\input{text/stability.fig}

\begin{document}


\begin{titlepage}

\begin{tabular}{l}
\noindent\DATE
\end{tabular}
\hfill
\begin{tabular}{l}
\PPrtNo
\end{tabular}

\vspace{1cm}

\begin{center}
\renewcommand{\thefootnote}{\fnsymbol{footnote}}
{
\LARGE \TITLE
}

\vspace{1.25cm}
{\large  \AUTHORS}

\vspace{1.25cm}

\INST
\end{center}

\vfill

\ABSTRACT                 

\vfill

\newpage
\end{titlepage}

\tableofcontents  \newpage

\section{Introduction}

A critical need for progress in high energy physics is the continued 
improvement of global QCD analysis to determine parton distribution functions
(PDFs), which link measured hadronic cross sections to the underlying partonic
processes of the fundamental theory. Precision tests of the Standard Model and
searches for New Physics in the next generation of collider programs at the
Tevatron and the LHC will depend on accurate PDFs and reliable estimates of
their uncertainties.

The vast majority of work on the analysis of PDFs and their application
to calculations of high-energy processes has been
performed at the next-to-leading order (NLO) approximation of perturbation
theory, i.e., 1-loop hard cross sections and 2-loop evolution kernels. For NLO
calculations in QCD, the order of magnitude of the neglected remainder terms
in the perturbative expansion is $\sim \! \alpha_{s}^{2}$ with respect to the
leading terms. Thus, the theoretical uncertainty of the predicted cross
sections at the energy scales of the colliders is expected to be on the 
order of a few percent.\footnote{%
Exceptions include specific processes for which the
perturbative expansion is known to converge more slowly
(e.g.~direct photon production); and processes near kinematic
boundaries, where resummation of large logarithms becomes
necessary (e.g.~small $x$ in DIS). These exceptional cases
have not so far become phenomenologically significant in global QCD
analysis.}
This level of accuracy is adequate for current phenomenology, since
experimental errors are generally comparable in size (for deep
inelastic scattering (DIS) measurements)
or larger (for most other processes).

In recent years, some preliminary next-to-next-leading-order (NNLO) analyses
have been carried out either for DIS alone \cite{Alekhin}, or in a global 
analysis context \cite{Mrstnnlo} (even if the necessary hard cross sections
for some processes, such as inclusive jet production, are not
yet available at this order).\footnote{%
The NNLO evolution kernel was also only known approximately at the time of
these analyses; but that gap has since been closed \cite{NnloKernel}.}
The differences with respect to the corresponding NLO analyses were 
indeed of the expected order of magnitude, including the expected somewhat 
larger differences with respect to power-counting in $\alpha_s$ that appear 
close to kinematic boundaries.

All other considerations being equal, a global analysis at NNLO must be
expected to have a higher accuracy. However,
NLO analyses can be adequate as long as their accuracy is
sufficient for the task, and as long as their predictions are stable with
respect to certain choices inherent in the analysis. Examples of those
choices are the functional forms used to parametrize the initial
nonperturbative parton distribution functions, and the selection of
experimental data sets included in the fit---along with the kinematic
cuts that are imposed on that data.

In global QCD analyses, kinematic cuts on the variables $x$, $Q$, $W$,
$p_{t}$, etc., are made in order to suppress higher-order contributions,
unaccounted edge-of-phase-space effects, power-law corrections, small-$x$
evolution effects, and other nonperturbative effects. In the absence of a
complete understanding of these effects, the optimal choice for the 
kinematic cuts must be determined empirically. We do so by varying the 
cuts and finding regions of
stability (i.e., internal consistency). Based on past
studies of this kind, CTEQ global analyses have adopted the following
``standard cuts'' for DIS data: $Q>2 \, \mathrm{GeV}$ and $W>3.5\,
\mathrm{GeV}$. The standard MRST analyses use cuts of $Q>1.41 \,
\mathrm{GeV}$ and $W>3.54 \, \mathrm{GeV}$.

The stability of NLO global analysis has, however, been seriously challenged
by recent MRST analyses, particularly \cite{mrst03} which found a 20\%
variation in the cross section predicted for $W$ production at the LHC---a
very important ``standard candle'' process for hadron colliders---when
certain cuts on input data are varied. If this instability is verified, it
would significantly impact the phenomenology of a wide range of physical
processes for the Tevatron Run II and the LHC. We have therefore performed an
independent study of this issue within the CTEQ global analysis framework.  
In addition, to explore the dependence of the results on
assumptions about the parametrization of PDFs at our starting scale
$Q_{0} \! = \! 1.3\, \mathrm{GeV}$, we have also studied the effect of 
allowing a negative gluon distribution at small $x$---a possibility which 
is favored by the MRST NLO analysis, and which appears to be tied to the 
stability issue.

In Sec.~2 we discuss issues relevant to the stability problem.
In Sec.~3 we summarize the theoretical and experimental inputs to the global
analyses.
In Sec.~4 we describe the detailed results of our study. The main finding
is that both the NLO PDFs and their physical predictions at the
Tevatron and the LHC are quite stable with respect to variations of the
cuts and the parametrization. Since this conclusion is quite
different from that of the MRST study, potential sources of the difference
are analyzed.
In Sec.~5, the prediction and uncertainty for $W$ production at the LHC 
are studied in more detail using the robust Lagrange Multiplier method, 
with particular attention to the stability issue.
Three Appendices contain more detailed discussions of three issues that arise
in the comparison of CTEQ and MRST analyses described in Sections 4 and 5:
(A) the definition of $\alpha_s$;
(B) the small-$x$ behavior of PDFs, including negative $g(x,Q)$; and
(C) the large-$x$ behavior of $g(x,Q)$ and spectator counting rules.

In addition to the CTEQ and MRST analyses, there are some other PDF
analysis efforts, which focus mainly on DIS data \cite{Alekhin,zeus,h1,gkk}.
However, these do not address the stability issue, because it is the
interplay between DIS, Drell-Yan (DY) and Jet data sets that raises that issue. 
For this reason, our discussions will consider only results from the two global
analysis groups that make use of all three types of hard processes.

\section{Issues related to the stability of NLO global analysis}
\label{sec:issues}

In this section we provide some background on the stability issue, before
describing our independent study of it in later sections.

The main evidence for instability of the NLO global analysis observed in
Ref.~\cite{mrst03} is shown in Fig.~\ref{fig:mrst}. 
Figure~\ref{fig:mrst}(a) shows the
variation of the predicted total cross section for $W$ production at the LHC,
as a function of the kinematic cut on the Bjorken variable $x$ in DIS. For the
largest $x$ cut ($0.01$) the NLO prediction is 20\% lower than the standard
prediction: the two predictions are clearly incompatible.
\figMrst
Figure~\ref{fig:mrst}(b) 
shows the predicted rapidity distributions for the $W$ boson.
The prediction from the ``conservative fit'' (the one with the largest 
$x$ cut) drops steeply compared to that of the standard fit outside the
central rapidity region, thus creating the drop in $\sigma_W$ seen in 
Fig.~\ref{fig:mrst}(a). This effect was attributed in Ref.~\cite{mrst03} 
to a ``tension'' between the Tevatron
inclusive jet data and the DIS data at small $x$ (HERA) and medium $x$ (NMC).
That tension is gradually relaxed as the $x$ cut is
raised, i.e., as more small-$x$ data are excluded. Evidently, removing the
HERA constraint, and thus effectively placing more emphasis on the inclusive
jet data, significantly changes the PDFs and the resulting prediction for
$\sigma_{W}$. In the MRST analysis, the combined data also pull
the gluon distribution to negative values at small $x$ and small $Q$. It is
likely that these two problems---the instability of the prediction on
$\sigma_{W}$ and the negative gluon PDF---are interrelated.

The CTEQ and MRST analyses use largely the same data sets, theory input
and methodology. Hence they usually yield results that are in general
agreement. However, minor differences between the choices made for
these inputs can, under some circumstances, give rise to significant
differences in the resulting PDFs and their predictions. For example, in
Fig.~\ref{fig:PdfDif}(a) the fractional uncertainty of the $u$ quark
distribution at $Q^{2}=10 \, \mathrm{GeV}^{2}$, normalized to 
CTEQ6.1M \cite{cteqjet},
is shown as the shaded band for $10^{-4}<x<0.9$. The comparison curves are
CTEQ6M \cite{cteq6}, MRST2002 \cite{mrst02}, 
MRST2003c \cite{mrst03}, and the reference CTEQ6.1M (horizontal
line).
We see that: (i) the uncertainty is small, $\sim \! 5\%$ for much of the
$x$ range; and (ii) with the exception of MRST2003c (``c'' for
\emph{conservative}) at small $x$, the fits agree reasonably
well. This reflects the tight constraints imposed mainly by the precise 
DIS and DY data.
\figPdfDif

The corresponding comparison for the gluon distribution is shown in
Fig.~\ref{fig:PdfDif}(b). We see that the fractional uncertainty is much 
larger in this case, especially for $x>0.25$. Even taking into account the 
size of the uncertainties, a difference in shape between the MRST and CTEQ 
gluon distributions over the full range of $x$ is evident. The differences 
between the MRST and CTEQ standard fits indicate how the small-to-medium-$x$
DIS data and the medium-to-large-$x$ Tevatron inclusive jet data are being fit
differently in the two analyses (to be discussed below). Also noticeable is
the change in shape of $g(x,Q)$ below $x \sim 0.1$ between the default and
conservative MRST distributions.  This difference shows the effect of 
cutting out DIS data at small $x$.

The gluon distribution is closely tied to the jet data. The stronger
gluon at high $x$ in CTEQ6.1 leads to a larger predicted cross
section at high jet $E_{T}$, in better agreement with the Tevatron data.
Quantitatively, $\chi^{2}$ for the Tevatron Run 1 CDF and D0 jet data
is 118 for 123 data points in the CTEQ6.1M fit. The corresponding numbers
for MRST2002 are $\sim \! 160$ for 115 data points \cite{mrst02,mrst03}.
Figure~\ref{fig:d0jet} shows four $\eta$ bins of the D0 data, along with the
theoretical curves obtained with CTEQ6.1M,
MRST2002 and MRST2003c PDFs.%
\footnote{The D0 data separated in $\eta$ bins are more sensitive to the
behavior of the gluon distribution over a wider range of $x$ than the CDF
data, which are limited to central rapidity. The highest $\eta$ bin measured
by D0 is not shown here since it is not included in the MRST analyses. We
thank Robert Thorne for providing the theoretical values of the MRST curves.} %
\figD
In the PDF parametrization adopted in the standard MRST fits, the rather high
value of $\chi^{2}$ for the jet cross section results from a trade-off with
the $\chi^{2}$ of DIS data at small-to-medium $x$ (hence the \emph{tension})
\cite{mrst02,mrst03}. This tension is relaxed only when DIS data with
$x<0.005$, and $Q^{2}<10 \, \mathrm{GeV}^2$ are removed from the fit,
resulting in the conservative fit MRST2003c, which reduces $\chi^2$ for
the jet data sets significantly \cite{mrst03,mrst04}. The reduction in
$\chi^2$ occurs mostly in the low transverse momentum range 
($100 - 200 \, \mathrm{GeV/c}$)
for the lower rapidity bins: it results from an interaction between the 
change in the predicted jet cross section and the shapes of the experimental
correlated systematic errors. These differences between the two fits are
attributable mainly to differences in the gluon distribution.  In contrast, in
the CTEQ6.1 fits, the already good $\chi^2$ for the jet data does not improve
noticeably when similar cuts are made (cf.~Sec.~\ref{sec:results}).

The significant differences between the MRST standard and conservative fits,
and their physical predictions (cf.~Fig.~\ref{fig:mrst}), highlight the
instability of these NLO QCD global analyses. The ``conservative'' fit,
although free from apparent tension, is not to be considered a serious
candidate for calculating safe physical predictions \cite{mrst03}.%
\footnote{We thank Robert Thorne for emphasizing this point 
(private communication).} %
First, the removal of the high precision small-$x$ and low-$Q$ DIS data
results in the loss of powerful constraints on the PDFs.
Therefore the uncertainty is increased for
physical predictions that depend on small-$x$ PDFs, which includes much of the
physics at the LHC. Second, in the particular case of MRST2003c,
the gluon distribution becomes strongly negative at small $x$ as seen in
Fig.~\ref{fig:ALefigquark4xx}(a) of Appendix \ref{app:NegPdf}.
Therefore unphysical negative predictions result for some quantities, such
as $F_L$ in DIS, and $d\sigma_W/dy$ at large $y$ for very high energies.

In the MRST analyses, NLO fits are unstable due to tension between the inclusive
jet data and the DIS data.
On the other hand, the CTEQ NLO global analyses do not show this
tension. It is therefore
important to investigate the stability issue in more detail within the CTEQ
framework, to determine whether the NLO analysis is viable.

\section{Inputs to the current analysis}
\label{sec:inputs}

The new global analyses in our stability study are extensions of the CTEQ6 
analysis.
We briefly summarize the theoretical and experimental inputs in this section.
Some of these features are relevant for later discussions
on the comparison of our results with those of Ref.~\cite{mrst03}.
For details, see the CTEQ6 paper \cite{cteq6}.

We use the {$\overline{\mathrm{MS}}$} scheme in the conventional PQCD
framework, with three light quark flavors ($u,d,s$).  The charm and bottom
partons are turned on above momentum scales ($\mu _{f}=Q$) equal to the heavy 
quark masses $m_{c}=1.3 \, \mathrm{GeV}$ and $m_{b}=4.5 \, \mathrm{GeV}$. 
To be consistent with the 
most common applications of PDFs in collider phenomenology, 
each parton flavor is treated as massless above its mass threshold.

The input nonperturbative PDFs are defined at an initial scale 
$\mu_{f}=Q_{0}=1.3 \, \mathrm{GeV}\ (=m_c)$ using functional forms that meet
certain criteria: 
(i) they must reflect qualitative physical behaviors
expected at small $x$ (Regge behavior) and large $x$ 
(spectator counting rules); and 
(ii) they must be flexible enough to allow for unknown nonperturbative 
behavior (to be determined by fitting data)\footnote{%
Unnecessarily restrictive parametrizations, which introduce artificial
correlations between the behavior of PDFs in different regions of the $x$
range, have been responsible for several wrong conclusions in past global QCD
analyses.}; while 
(iii) they should not involve more free parameters than can
be constrained by available data.  In general, we use the functional form
\begin{eqnarray}
xf(x,Q_{0}) = A_{0} \, x^{A_{1}} \, (1-x)^{A_{2}} \, \mathrm{e}^{A_{3} \, x} 
\, (1+\mathrm{e}^{A_{4}}\,x)^{A_{5}}
\end{eqnarray}
for each flavor (see \cite{cteq6} for motivation and explanation).
This generic form is modified as necessary to study specific 
issues---such as whether a negative gluon distribution at small $x$ is 
indicated by data.

The experimental data sets that are used in the new analyses are essentially
the same as those of CTEQ6, with minor updates.\footnote{%
For instance, a third set of H1 data \cite{H1new} has been added to the two
sets used in \cite{cteq6}.} As mentioned in the previous sections, 
kinematic cuts on the input data sets are systematically varied as a part of
the stability study.

\section{Results on the stability of NLO global analysis}
\label{sec:results}

The stability of our NLO global analysis is investigated by
varying the inherent choices that must be made to perform the analysis.
These choices include the selection of experimental data points based on
kinematic cuts, the functional forms used to parametrize the initial
nonperturbative parton distribution functions, and the treatment of
$\alpha_s$.  Sections \ref{subsec:cuts}--\ref{subsec:Xsec} discuss the 
kinematic cuts and the form of
the gluon distribution, which relate directly to the ``tension'' found
in \cite{mrst03} that motivated this study.
Section \ref{subsec:alphas} discusses the role of different assumptions on
$\alpha_s(Q)$.

The stability of the results is most conveniently measured by differences in
the global $\chi^2$ for the relevant fits. To quantitatively define a change
of $\chi^{2}$ that characterizes a significant change in the quality of the
PDF fit is a difficult issue in global QCD analysis. In the context of the
current analysis, we have argued that an increase by
$\Delta\chi^{2}\sim 100$ (for $\sim$ \! 2000 data points) represents roughly
a 90\% confidence level uncertainty on PDFs due to the uncertainties of the
current input experimental data \cite{MultiVar,LM,Hesse,cteq6}. In other
words, PDFs with $\chi^2 - \chi^2_{\mathrm{Best Fit}} > 100$ are regarded as
not tolerated by current data. This {\em tolerance} will provide a useful
yardstick for judging the significance of the fits in our stability study,
because currently available
experimental data cannot distinguish much finer differences.\footnote{%
In the future, when systematic errors in key experiments are reduced and when
different experimental data sets become more compatible with each other, this
tolerance measure will shrink in size.}

\subsection{Stability of global fits: Kinematic cuts on input data }
\label{subsec:cuts}

The CTEQ6 and previous CTEQ global fits imposed ``standard'' cuts $Q > 2 \,
\mathrm{GeV}$ and $W > 3.5 \, \mathrm{GeV}$ on the input data set, in order
to suppress higher-order terms in the perturbative expansion and the effects
of resummation and power-law (``higher twist'') corrections. We examine in
this section the effect of stronger cuts on $Q$ to see if the fits are
stable. We also examine the effect of imposing cuts on $x$, which should
serve to suppress any errors due to deviations from DGLAP evolution, such as
those predicted by BFKL. The idea is that any inconsistency in
the global fit due to data points near the boundary of the accepted region
will be revealed by an improvement in the fit to the data that remain after
those near-boundary points have been removed. In other words,
the decrease in $\chi^2$ for the subset of data that is retained, when the
PDF shape parameters are refitted to that subset alone, measures the degree
to which the fit to that subset was distorted in the original fit by
compromises imposed by the data at low $x$ and/or low $Q$.

The main results of this study are presented in Table~\ref{tab:tableI}. Three
fits are shown, from three choices of the exclusionary cuts on input data as
specified in the table. They are labeled `standard', `intermediate' and
`strong'. $N_{\rm pts}$ is the number of data points that pass the cuts in each
case, and
$\chi^2_{N_{\rm pts}}$ is the $\chi^2$ value for that subset of data. The fact
that the changes in $\chi^2$ in each column are insignificant compared
to the uncertainty tolerance is strong evidence that our NLO global fit
results are very stable with respect
to choices of kinematic cuts. %
\tblI
As an example, note that the subset of 1588 data points that pass the 
\textit{strong} cuts ($Q > 3.162 \, \mathrm{GeV}$ and $x > 0.005$) are fit with
$\chi^2 \! =  \! 1573$ when fitted by themselves; whereas the compromises that 
are needed to fit the full standard data set force $\chi^2$ for this subset up 
to $1583$.  This small increase---only $10$ in the total $\chi^2$ for this 
large subset---is an order of magnitude smaller than the increase that would 
represent a significant change in the quality of the fit according to our 
tolerance criterion for uncertainties.

\subsection{Stability of global fits: negative gluon at small $x$?}
\label{subsec:NegGlu}

We have extended the analysis to a series of fits in which the gluon
distribution $g(x)$ is allowed to be negative at small $x$, at the scale
$Q_0 \! = \! 1.3 \, \mathrm{GeV}$ where we begin the DGLAP 
evolution.\footnote{%
To allow $g(x,Q_0)<0$, we include a factor $(1 + a x^b)$ in it, where $a<0$
and $b<0$ are allowed.  We have checked that this form can accurately 
mimic the form used in MRST2003c.}
The purpose of this
additional study is to determine whether the feature of a negative gluon PDF
is a key element in the stability puzzle, as suggested by the findings of
\cite{mrst03}. The results are presented in Table~\ref{tab:tableII}. Even in
this extended case, we find no evidence of instability. For example, $\chi^2$ 
for the
subset of 1588 points that pass the \textit{strong} cuts
increases only from 1570 to 1579 when the fit is extended to include
the full standard data set.

\tblII

Comparing the elements of Table~\ref{tab:tableI} and Table~\ref{tab:tableII}
shows that our fits with $g(x) < 0$ have slightly smaller values of
$\chi^2$: e.g., $2011$ versus $2023$ for the standard cuts.
However, the difference $\Delta \chi^2 \! = \! 12$ between these values is 
again not significant according to our tolerance criterion.

Negative parton distributions in a given renormalization and factorization
scheme are not strictly forbidden by theory. However, a PDF set that leads to
any negative cross sections---either of practical importance or as a matter of
principle---must be regarded as unphysical. Therefore to establish the 
viability of a PDF
set with negative PDFs is very difficult: the negative PDFs
can be enhanced in a special kinematic region for
a specific cross section, leading to a negative cross section.\footnote{%
For instance, the negative gluon distributions of \cite{mrst03}
give rise to negative $F_L$ at low $x$ and negative $d\sigma_W/dy$
at large $|y|$ and high $E$.}

Our results from a parametrization that allows $g(x)<0$ lead us to conclude
that a negative gluon distribution may be permitted, but is certainly not 
mandated, by our analysis.
Further discussion of this point, with specific examples from our fits and
those of \cite{mrst03}, is contained in Appendix \ref{app:NegPdf}.

\subsection{Stability of physical predictions}
\label{subsec:Xsec}
The last columns of Tables~\ref{tab:tableI} and \ref{tab:tableII} show the 
predicted cross section for
$W^+ + W^-$ production at the LHC. This prediction is also very stable: it
changes by only $1.6 \%$ for the positive-definite gluon parametrization,
which is substantially less than the overall PDF uncertainty of $\sigma_W$
estimated previously with the standard cuts.  For the negative gluon 
parametrization, the change is $4 \%$---larger, but still less than the 
overall PDF uncertainty.
These results are explicitly displayed, and 
compared to the MRST results of Fig.~\ref{fig:mrst}, in Fig.~\ref{fig:WtotXs}.
\figWtotXs
We see that this physical prediction is indeed insensitive to the kinematic
cuts used for the fits, and to the assumption on the positive definiteness
of the gluon distribution.
We have obtained similar results (not shown) for the individual $W^+$ and 
$W^-$ cross sections at the LHC and at the Tevatron.
A more focused study on the uncertainty of the LHC prediction and its variation
with the kinematic cuts, using the Lagrange Multiplier method,
is given in Sec.~\ref{sec:WtotXs}.

This section has demonstrated stability of our NLO global fits with respect 
to cuts on $x$ and $Q$ and with respect to the parametrization of the gluon 
input.  This result is consistent with the expected numerical accuracy of the 
PQCD expansion.  However, it is in apparent disagreement with the findings 
of \cite{mrst03}. The two analyses have some other differences, including 
the treatment of $\alpha_s$ which we examine next.

\subsection{Stability and $\alpha_s$}
\label{subsec:alphas}

The CTEQ5/6/6.1 PDF sets were extracted assuming 
$\alpha_s(m_Z) = 0.118\,$.
This value was chosen to approximate the world average, thereby to
incorporate the rather strong constraints from measurements---especially 
from LEP---that are not otherwise included in our input data set.
To examine the influence of $\alpha_s(m_Z)$ on the quality and stability
of the fit, we have made a series of fits with different choices for
$\alpha_s(m_Z)$.
This exercise provides a further test of the reliability of our analysis.

The results for $\chi^2$ as a function of $\alpha_s(m_Z)$ take the parabolic
form shown as the solid curve in Fig.~\ref{fig:ALmfiggluon12y}(a).
\figALm
Qualitatively, the value of $\alpha_s(m_Z)$ preferred by the global
fit is in reasonable agreement with the World Average, which lends support
to the idea that NLO QCD is working successfully in the global fit.
To obtain a quantitative result, we assume that $\Delta \chi^2 = 37$ 
defines a $1\, \sigma$ error (based on the estimated 90\% C.L. range for 
$\Delta \chi^2 = 100$ mentioned previously).  In this way we obtain
\begin{eqnarray}
\alpha_s(m_Z) =  0.1169 \pm 0.0045 \; (0.1148 \pm 0.0050)
\label{eq:alphastandardcuts}
\end{eqnarray}
from the PDF fit using the \textit{standard} data cuts, 
assuming $g(x) \! > \! 0$ ($g(x) \! < \! 0$ allowed).  
These results are fully consistent with the current world average
$0.1187 \pm 0.0020$\cite{PDG}, or with the LEP QCD working group average
$0.1201 \pm 0.0003 \pm 0.0048$.
This consistency lends confidence to the standard
analysis---including the ``standard'' data cuts used in it.
The average $\chi^2$ per data point at the minimum,
$\chi^2/N = 2020 / 1926 = 1.049$ for the standard cut fit, is also
comfortably close to $1$.

The solid curve in Fig.~\ref{fig:ALmfiggluon12y}(b)
shows the effect of imposing the ``strong'' data cuts
($Q > 3.162 \, \mathrm{GeV}$, $x > 0.005$).
The allowed range in $\chi^2$ should scale as the number of data points,
so we estimate the $1 \, \sigma$ uncertainty range in the case of the
strong cuts as the range over which $\chi^2$ increases by
$37 \times (1588/1926)$
above its minimum value.
This leads to 
\begin{eqnarray}
\alpha_s(m_Z) =  0.1168 \pm 0.0044 \; (0.1159 \pm 0.0051)
\label{eq:alphastrongcuts}
\end{eqnarray}
from the PDF fit using the \textit{strong} data cuts, 
assuming $g(x) \! > \! 0$ ($g(x) \! < \! 0$ allowed).  
The similarity between this result and Eq.~(\ref{eq:alphastandardcuts})
shows that the fit is very stable with respect to the cuts.

The dotted curves in Fig.~\ref{fig:ALmfiggluon12y} show the effect of
allowing the gluon distribution $g(x,Q_0)$ to be negative at small $x$. 
The resulting uncertainties in $\alpha_s$ are somewhat larger than for the 
positive gluon cases. In addition, the minimum $\chi^2$ values are somewhat 
lower than for positive-definite gluons. However, in view of the larger number 
of degrees of freedom in the fit and the reservations expressed previously 
concerning negative parton distributions, we do not consider these differences
in minimum $\chi^2$ persuasive. 

In addition to the possible range of values of $\alpha_s(m_Z)$,
there is an uncertainty caused by ambiguity in how to define
$\alpha_s(Q)$ at NLO.
We show in Appendix \ref{app:alphas} that this ambiguity also has little
effect on the results of the global fits or their stability.


\subsection{Comments and Discussion}
\label{subsec:compare}

The results for $\chi^2_{1770}$ and $\chi^2_{1588}$ in 
Tables~\ref{tab:tableI} and \ref{tab:tableII}, and the results on
$\sigma_W^{\mathrm{LHC}}$ given in those Tables and in Fig.~\ref{fig:WtotXs}
show a reassuring stability of the global fits.
This confirms the general expectations for the PQCD expansion, and
lends confidence to the extensive body of NLO phenomenological work that is 
being done in connection with current and future collider physics programs.
However, it is important to ask why our results differ from those of
\cite{mrst03}. Two separate issues are involved in the comparison between the
two global analysis programs.

First, the instability of the NLO analysis observed in \cite{mrst03} appeared
originally to result from a ``tension'' between the Tevatron inclusive jet
data (mostly at medium and large $x$) and the DIS data at small 
and medium $x$. This tension has been a persistent feature of 
recent MRST analyses. However, CTEQ analyses, including the current study,
have consistently not seen it.  The difference appears to be due to
the behavior of the gluon distribution at large $x$. The CTEQ input
gluon distribution is consistently higher in the large $x$ region, which
produces a much better fit to the CDF and D0 jet production cross section
without affecting the fits to the DIS data. This point has been confirmed
recently by a new MRST paper \cite{mrst04}, which uses an input gluon
distribution quite similar in shape to the CTEQ $g(x)$.  The
large-$x$ behavior of the relevant PDF sets in discussed in 
Appendix \ref{app:largex}.

The second issue concerns negative gluons at small $x$.\footnote{%
Although initially thought to be related to the large-$x$ issue
through momentum sum rule constraints, the connection is less clear now,
because of the advance in \cite{mrst04}.}
Whereas we find only marginal differences in the quality of the global fits
when the input gluon function is allowed to become negative,
MRST has found a strong pull toward negative gluon in their analyses.
Furthermore, their gluon distribution becomes increasingly negative
as the $x$ cut is raised.
The increasingly negative gluon distribution at small $x$,
through its influence on the sea quark distributions
via QCD evolution, is directly responsible for the rapid decrease of
$d\sigma_{W}^{\mathrm{LHC}}/dy$ outside the central rapidity
region, and the consequent decrease of the total $\sigma_W^{\mathrm{LHC}}$,
as seen in Fig.~\ref{fig:mrst}.
Further details on the small-$x$ behavior of the relevant PDFs are
discussed in Appendix \ref{app:NegPdf}.

The source of the different conclusions about a negative gluon PDF is
tentatively identified in Appendix \ref{app:NegPdf} as a difference in
assumptions about the input gluon distribution.
At any rate, because the improvement to the fit is small, and because of
the reservations expressed earlier about negative PDFs, we do not
believe that allowing negative gluons is necessary to the global analysis.


\section{Stability and Uncertainty of $\sigma_{W}$ at the LHC}
\label{sec:WtotXs}

In this section, we study in detail the stability of the NLO prediction
for the cross section $\sigma_{W}$ for $W^+ + W^-$ production
at the LHC, using the Lagrange Multiplier (LM) method of
Refs.~\cite{MultiVar,LM,Hesse}.
Specifically, we perform a series of
fits to the global data set that are constrained to specific values of
$\sigma_W$ close to the best-fit prediction.
The resulting variation of $\chi^{2}$ versus $\sigma_{W}$ measures
the uncertainty of the prediction.
We repeat the constrained fits for each case of fitting choices
(parametrization and kinematic cuts).  In this way we gain an 
understanding of the stability of the uncertainty, in addition 
to the stability of the central prediction.  
The LM method is more 
robust than simply using the 40 eigenvector sets from CTEQ6.1, 
which were obtained using the Hessian method, because the LM 
technique probes the full parameter space instead of only the 
subspace of 20 free parameters that were varied in the Hessian 
method.  We find that the uncertainty range obtained by the LM 
method is not much different from that obtained from the 40 sets,
which demonstrates that the Hessian estimate of the uncertainty 
is not biased by choices made in the parametrizations at $Q_0$.

Figure~\ref{fig:ChiVsSigCutsPG} shows the results of the
LM study for the three sets of kinematic cuts described in 
Table~\ref{tab:tableI}, all of which have a positive-definite gluon 
distribution.
\figChiVsSigCutsPG%
The $\chi^2$ shown along the vertical axis is normalized to its value for the
best fit in each series.%
\footnote{ Neither the absolute value of $\chi^2$, nor its increment above
the respective minimum, are suitable for comparison, because the different
cuts make the number of data points quite different for the three cases.} %
In all three series, $\chi^2$ depends almost quadratically on
$\sigma_W$. We observe several features:
\begin{Simlis}{1 em}
{\item The location of the minimum of each curve represents the best-fit
prediction for $\sigma_W^{\mathrm{LHC}}$ for the corresponding choice of
exclusionary cuts. The fact that the three minima are close together displays
the stability of the predicted cross section already seen in 
Table~\ref{tab:tableI}.

\item
Although more restrictive cuts make the global fit less sensitive to possible
contributions from resummation, power-law and other nonperturbative effects,
the loss of constraints caused by the removal of precision HERA data points 
at small $x$ and low $Q$ results directly in increased
uncertainties on the PDF parameters and their physical predictions.
This is shown in Fig.~\ref{fig:ChiVsSigCutsPG}
by the increase of the width of the curves with stronger cuts.
The uncertainty of the predicted $\sigma_W$ increases by more than a factor 
of 2 in going from the standard cuts to the strong cuts.

\item
The uncertainty range for $\sigma_W$ calculated from the 40 eigenvector 
uncertainty sets of CTEQ6.1 is $\pm 5.5\,\%$.  The width of the $\chi^2$ 
parabola in Fig.~\ref{fig:ChiVsSigCutsPG} at $\Delta\chi^2=100$ for the 
standard cuts is similar, though it is slightly larger
because the experimental normalizations, and all of the PDF shape parameters 
rather than just 20 of them, are treated as free in the LM fits.  The LM 
method thus confirms that the estimate based on the eigenvector sets is 
reasonably good.
}
\end{Simlis}

Figure~\ref{fig:ChiVsSigCutsNG}(a) shows the results for
the cases of standard/intermediate/strong cuts summarized in 
Table~\ref{tab:tableII} when the
gluon distribution is allowed to be negative at small $Q$ and $x$.
\figChiVsSigCutsNG
In this case, we observe:

\begin{Simlis}{1 em}
{
\item
The stability of the best fits, represented by the minima of the curves,
is again apparent.

\item
With strong cuts and allowing negative gluons,
the uncertainty range of $\sigma_{W}$ expands considerably,
especially toward low values of $\sigma_{W}$.
The solutions at the extreme low end of the $\sigma_{W}$
range are most likely unphysical, since a strongly 
negative gluon distribution at small $x$ and $Q$ can drive the
quark distributions negative at $x \sim 10^{-4}$ at moderate values
of $Q$ by QCD evolution (cf.~Appendix \ref{app:NegPdf}).
}
\end{Simlis}

Figure~\ref{fig:ChiVsSigCutsNG}(b) compares
the two LM series obtained using the standard cuts,
with and without the positive definiteness requirement.
We observe:

\begin{Simlis}{1 em}
{
\item
Removing the positive definiteness condition necessarily lowers the value of
$\chi^2$, because more possibilities are opened up in
the $\chi^2$ minimization procedure.  But the decrease is insignificant
compared to other sources of uncertainty.
Thus, a negative gluon PDF is allowed, but not required.

\item The minima of the two curves occur at approximately the same
$\sigma_{W}$. Allowing a negative gluon makes no
significant change in the central prediction---merely a
decrease of about $1\,\%$, which is small compared to the 
overall PDF uncertainty. 

\item
For the standard set of cuts, allowing a negative gluon PDF would expand 
the uncertainty range only slightly.  

\item
The dot-dash curve in Fig.~\ref{fig:ChiVsSigCutsNG}(b) is obtained by 
modifying the gluon parametrization at $Q_0$ by including a factor 
$e^{-(a/x)^b}$, which can suppress $g(x,Q_0)$ without allowing it to 
become negative.  It demonstrates that most, if not all, of the reduction
in $\chi^2$ obtainable by a low-$x$ suppression does not require $g(x)$
to be negative.  We do not, however, find this small reduction 
in $\chi^2$ persuasive enough to give up the assumption of approximate 
Regge behavior at small $x$.
}
\end{Simlis}


\section{Conclusions}

Motivated by its importance to all aspects of collider physics phenomenology
at the Tevatron and the LHC, we have examined the stability of the NLO QCD
global analysis with respect to certain variations in its input, in
particular, the selection of input experimental data and the functional form
of the nonperturbative gluon distribution.

As increasingly stringent kinematic cuts at higher $x$ and $Q$ are placed on
the input data, in order to exclude potentially unsafe regions
of phase space, we find no significant improvement in the quality of the
fit, as measured by the $\chi^2$ of the retained data. In particular, we
do not observe the ``tension'' discussed in recent MRST analyses.
Simultaneous good fits to the HERA and Tevatron jet data are obtained for the
full range of cuts explored. Predictions for the $W$ cross section at both the
Tevatron and LHC were examined. As data at lower $x$ and $Q$ are removed
from the analysis, the central value remains quite stable, while the
uncertainty on the predicted $\sigma_{W}$ increases.  

We have repeated this analysis with the gluon distribution allowed to
assume negative values at small $x$.   
There is a slight reduction in the global $\chi^2$
(as is expected whenever the fitting parameter space is expanded), but the
size of the reduction is too small to be of 
physical significance according to our analysis framework.
As data at lower $x$ and $Q$ are removed, the central prediction for 
$\sigma_{W}$ again remains quite stable, while its uncertainty increases.  
In this case, when ``strong'' cuts in  $x$ and $Q$ are made, similar to those 
of the MRST2003c ``conservative'' fit, the allowed range for $\sigma_W$ expands 
far enough to include the MRST2003c prediction.  
However it is not the central prediction, and we find no 
evidence of a tension between data sets that would support requiring such 
a strong cut; or support allowing a negative gluon distribution at small 
$x$.  Thus we conclude that the predictions for $\sigma_W$ are sufficiently 
well defined that its use as luminosity monitor for QCD processes 
(``standard candle'') at the Tevatron and LHC is viable after all.

We have examined a number of aspects of our analysis that might account 
for the difference between our stability study and that of \cite{mrst03}.
A likely candidate seems to be that in order to obtain stability, it is 
necessary to allow a rather free parametrization of the input gluon 
distribution.  This suspicion is seconded by recent work by MRST 
\cite{mrst04}, in which a different gluon parametrization appears to 
lead to a best-fit gluon distribution that is close to that of CTEQ6.

In summary, we have found that the NLO PDFs and their physical predictions at
the Tevatron and LHC are quite stable with respect to variations of the
kinematic cuts and the PDF parametrization after all. Thus, the NLO framework 
should provide sufficient accuracy for phenomenology at both Run II of the 
Tevatron and at the LHC. Further improvement will be possible with a NNLO 
global QCD analysis. A fully global analysis at NNLO, however, must wait for 
the completion of the NNLO QCD calculation of all of the relevant hard 
processes---in particular the inclusive jet cross section.

\paragraph{Acknowledgements:}
We thank James Stirling and Robert Thorne for many informative
and stimulating communications concerning the similarities and differences of
the MRST and CTEQ analyses. This research is supported by the National Science
Foundation.



\appendix

\section{Definition of $\protect\alpha_s(Q)$ at NLO}

\label{app:alphas}

A subtle difference between various NLO global analyses arises from the choice
of definition for the variation of $\alpha_s(Q)$ at NLO.  The various choices
differ only at NNLO, so \textit{a priori} they are equally valid at NLO.  
Principal definitions in use are
\begin{enumerate}
\item
Exact solution of the truncated renormalization group equation:
\begin{eqnarray}
\mu \, d\alpha /d\mu =c_{1}\alpha^2\,+\,c_2\alpha^3 \; ,
\end{eqnarray}%
where $c_1 = -\beta_0/2\pi$ with $\beta_0=11-(2/3)\mathrm{n}_f$, and $c_2 =
-\beta_1/8\pi^2$ with $\beta_1=102-(38/3)\mathrm{n}_f$. This is the recipe
used in the QCD evolution program QCDNUM \cite{QcdNum}, which is used by
several groups, including ZEUS.

\item
The original $\overline{\mathrm{MS}}$ definition at NLO \cite{MsbAlf}:
\begin{eqnarray}
\alpha (Q)=c_{3}\,[1-c_{4}\ln (L)/L]/L \; ,
\end{eqnarray}
where
$L = \ln (Q^2 /\Lambda^2)$,
$c_3 = -2/c_1$, and
$c_4 = -2\,c_2/c_1^2 \,$.
This is the standard definition used in CTEQ global analyses.

\item
The form chosen by MRST is less simple to state, since for $Q > m_b$ it contains
a NNLO term that depends on
$1/\alpha(m_b)-1/\alpha(m_c)$.  However, it is numerically very similar to the
QCDNUM choice.

\end{enumerate}

\figALb

Figure~\ref{fig:ALbfigmodes1a} shows that the MRST and QCDNUM forms are 
almost the same numerically; and that the difference between either of them 
and the CTEQ form is quite small in the region $Q > 2 \, \mathrm{GeV}$ where 
we fit data. In particular, that difference is much smaller than the 
difference caused by reasonable changes in  $\alpha_s(M_z)$.

\figALn

The dependence of $\chi^2$ for the global fit on $\alpha_s(m_Z)$ is shown
in Fig.~\ref{fig:ALnfigalpha12x} for \textit{standard} and
\textit{strong} cuts.
One sees that the choice of form for $\alpha_s(Q)$ has very little effect
on the quality of the fit, which is a satisfying indication of the
stability of the fit with respect to this arbitrary choice that must be
made to carry it out.  The similarity of the two figures shows that
the fit is stable with respect to kinematic cuts as well.

In detail, the two choices produce somewhat
different best-fit values for $\alpha_s(m_Z)$.  This can easily be
understood using Fig.~\ref{fig:ALbfigmodes1a}: for a given $\alpha_s(m_Z)$,
the QCDNUM choice gives a slightly smaller $\alpha_s(Q)$ in the region
of $Q$---mostly much smaller than $m_Z$---that is important in the fit.

The uncertainties of $\alpha_s(Q)$ lead to an uncertainty in the prediction
for $\sigma_W$ at the LHC.  In particular, the four fits with standard cuts,
which are shown in Fig.~\ref{fig:ALnfigalpha12x}(a), span a range of $\pm 1.5
\%$ ($\pm 2.9 \%$) in $\sigma_W$ for $\alpha_s(m_Z)=0.116$ -- $0.120$ ($0.114$
-- $0.122$). The four fits with \textit{strong} cuts, shown in
Fig.~\ref{fig:ALnfigalpha12x}(b), span a much larger range: $\pm 8.2 \%$ ($\pm
10.1 \%$) in $\sigma_W$. Once again, we see the loss of predictive power 
when too much data is removed from the input.

\section{PDFs at small $\protect x$: Do they go negative?}

\label{app:NegPdf}

As mentioned in the text, the behavior of the gluon distribution (and
through DGLAP evolution, the sea quark distributions) at small $x$ and
low $Q$ appears to be an open issue at the present time.  In particular,
there is a question as to whether the data allow or suggest that these
distributions become negative at small $x$.  We discuss the situation
in more detail in this appendix.

Figure~\ref{fig:ALnfigalpha12x} shows that allowing $g(x) < 0$ (by inserting a
factor $(1 + a x^b)$ into the standard CTEQ parametrization for $g(x)$ at $Q_0
= 1.3 \, \mathrm{GeV}$, with $a$ and $b$ allowed to be negative) leads to a
small improvement in the global fit: $\chi^2$ decreases by about $20$ (the
difference between the minima of the solid and dotted curves in 
Fig.~\ref{fig:ALnfigalpha12x}(a)). This decrease is well within the tolerance 
of our uncertainty range---especially in view of the fact that almost 
\emph{any} additional freedom in the fitting functions is expected to permit 
at least a small decrease in $\chi^2$.\footnote{%
For instance, suppressing the standard 
$g(x,Q_0)$ by a factor $e^{-(0.0011/x)^{1.67}}$ is sufficient to lower 
the best-fit $\chi^2$ by 10 units without making $g(x)$ negative; 
the resulting distribution is shown as the dot-dash curve in 
Fig.~\ref{fig:ALefigquark4xx}(a).
\label{fn:GluonSuppress}}
The change in $\chi^2$ for the case of the \textit{strong} data cuts 
(Fig.~\ref{fig:ALnfigalpha12x}(b)) is about $13$---again not persuasive.

The MRST2003c NLO fit has a much more negative gluon than any of the fits
described here, as shown in Fig.~\ref{fig:ALefigquark4xx}(a). 
Its negative region is so strong that it evolves to produce
negative sea quark distributions for $x < 2 \times 10^{-5}$ at $Q = 100 \,
\mathrm{GeV}$ (see Fig.~\ref{fig:ALefigquark4xx}(b)).  The suppression of these 
sea quark distributions at small $x$ near where they pass through zero is
responsible for the much smaller $\sigma_W$ ($16.12\,\mathrm{nb}$) predicted 
by the MRST2003c NLO PDFs, because $W$'s are produced at large rapidity 
by the annihilation of a quark at large $x$ with an antiquark at
very small $x$. The small $x$ values are well below the $x$ cut on
input data for the fit that created these PDFs, so the prediction is
intrinsically unreliable. In fact, at a slightly higher energy, say $\sqrt{s}
= 40 \, \mathrm{TeV}$, the same PDFs predict a negative \emph{cross section} 
over a substantial region of large rapidity.
\figALe

We are able to reproduce a similar suppression of the predicted $\sigma_W$ in
our fits that allow a negative $g(x)$ at $Q_0$ only by simultaneously (1)
imposing the \textit{strong} data cuts on $x$ and $Q$; and (2) increasing the
fit $\chi^2$ by $\sim \! 20$ units by employing the Lagrange Multiplier
method to force $\sigma_W$ downward.  This modest increase in
$\chi^2$ is acceptable according to our tolerance criterion.  However, 
the stability of
our fits with respect to the cuts makes it unnatural to impose the strong
cuts. And, as mentioned above, these PDFs should not be trusted in the first 
place, in the small-$x$ region that lies below the $x$ cut on the data input 
to the fit.

Figure~\ref{fig:ALefigquark4xx}(b) shows the $\bar{u}(x)$ distribution at $Q =
100 \, \mathrm{GeV}$.  (The $u(x)$, $d(x)$, $\bar{u}(x)$, $\bar{d}(x)$ are 
nearly identical at small $x$.)  The CTEQ6.1 and MRST2002 curves
are very similar, while MRST2003c turns negative at small $x$.
Our best fit with $g(x) < 0$ is quite similar to CTEQ6.1.  Even when
$\sigma_W$ is forced smaller by a Lagrange multiplier that raises $\chi^2$ by
$70$, the distribution is not greatly different.  Substantially
different behavior is obtained only when we impose the \textit{strong} data
cuts and force $\sigma_W$ small by a Lagrange multiplier 
(dot-dot-dash curve).

\section{Gluon distribution at large $\protect x$: Do counting rules count?}

\label{app:largex}

The behavior of the gluon distribution at large $x$ strongly affects 
the fit to inclusive jet production data. It therefore has a direct bearing on
whether the jet data can be described simultaneously with the precision DIS
data. 

Figure~\ref{fig:ALgfiggluon3x}(a) shows the gluon distribution at $Q = 100 \,
\mathrm{GeV}$ from various PDF fits.  The solutions that fit the jet data best
are those with a rather strong gluon at large $x$, such as 
CTEQ6.1 (the solid curve).  
The two dotted curves are eigenvector sets 29 and 30, which are the
members of the 40 eigenvector uncertainty PDF sets that have the most extreme
gluon distributions \cite{cteqjet}. MRST2002 lies just at the edge of this
range of uncertainty, which presumably accounts for the ``tension'' MRST find
between DIS and jets with this solution.   MRST2003c is slightly closer to
the CTEQ result, while the most recent MRST2004 is much closer to it.

\figALg

In all PDF analyses, the gluon distribution at $Q_0$ has been parametrized in
a form that varies as $(1-x)^a$ as $x \to 1$. We have treated $a$ as a free
parameter in the fitting, just like all of the other parton shape parameters
at $Q_0 = 1.3 \, \mathrm{GeV}$. Fig.~\ref{fig:ALgfiggluon3x}(b) shows the gluon
distributions as a function of $1-x$ on a log-log plot.  The approximately
straight-line behavior at small $1-x$ shows that an effective $(1-x)^a$
dependence survives the effects of the other parameters. From
the slopes of the straight lines, we find that the effective power $a$ is
about $1.7$ for CTEQ6.1, and varies from $0.8$ to $3.6$ over the 40
eigenvector sets.  This parameter is therefore not strongly constrained by 
the global fit; though it tends to be smaller than one would have expected 
on the basis of the ``spectator counting rules,'' \cite{counting rules} which
predict a faster fall-off for gluons than for valence quarks at $x \to 1$.

The MRST2003c gluon has $a \approx 3.5$, similar to the steepest fall-off 
of the 40 uncertainty sets of CTEQ6.1.  MRST2002 is even steeper, with 
$a \approx 4.1$.  This difference between the MRST and CTEQ fits does not 
result directly from an attempt to satisfy the counting rules, since the 
parameter $a$ is treated as free in both fits.  But the form of 
parametrization used in MRST2003c may tie the $x \to 1 $ behavior more 
closely to the behavior at intermediate $x$, and in that way not allow $a$ 
to come out small.

Theoretical constraints for the parametrization of nonperturbative input
parton distributions at small $x$ (Regge behavior) and large $x$ (spectator
counting rules) are, at best, only qualitative guides, since there is no
reason to impose the suggested behavior at any particular scale $Q_0$, nor for
PDFs in any particular factorization scheme. Constraints imposed with one
choice of scale and scheme can become rather different at another scale and
scheme. In particular, the power $a$ of the $(1-x)^a$ factor for the input
gluon distribution is well-known to be quite
sensitive to the choice of factorization scheme.\footnote{%
For an early discussion of this point, see the review 
article \cite{OwensTung}. \label{fn:OwensTung}}%
A recent paper by MRST \cite{mrst04} takes advantage of this feature, and
obtains a much better fit to the inclusive jet data with a gluon
parametrization in the DIS scheme that is close to the counting rule value
(and similar to what they had used before in the $\overline{\mathrm{MS}}$
scheme). The resulting $g(x,Q_0)$ in the $\overline{\mathrm{MS}}$ scheme, as
shown in Fig.~\ref{fig:ALgfiggluon3x}, turns out to be essentially similar to
that of the CTEQ analyses, which was arrived at in the global fit by an
unconstrained parametrization.  The fact that the mrst2002 fit, which used 
a supposedly unconstrained model for the input $g(x)$, was improved by 
imposing an additional condition on the large-$x$ behavior of $g(x)$ 
suggests that the original parametrization was not sufficiently flexible.

By contrast, it is interesting to note that the large-$x$ valence quark 
behavior is expected to be relatively insensitive to the choice of scheme 
(cf.~footnote \ref{fn:OwensTung}). Phenomenologically, the powers $a$ for 
both the $u_\mathrm{val}$ and $d_\mathrm{val}$ distributions are found to 
be quite stable, and their values, although slightly dependent on the choice 
of $Q_0$, are generally consistent with expectations from the counting 
rules. These results provide additional evidence that the underlying 
theoretical framework of global QCD analysis is physically sound.

\end{document}